# A validation of the short-form classroom community scale for undergraduate mathematics and statistics students

Maria Tackett[1], Shira Viel[2], Kim Manturuk[3]

## Abstract

This study examines Cho & Demmans Epp's short-form adaptation of Rovai's well-known Classroom Community Scale (CCS-SF) as a measure of classroom community among introductory undergraduate math and statistics students. A series of statistical analyses were conducted to investigate the validity of the CCS-SF for this new population. Data were collected from 351 students enrolled in 21 online classes, offered for credit in Fall 2020 and Spring 2021 at a private university in the United States. Further confirmatory analysis was conducted with data from 128 undergraduates enrolled in 13 in-person and hybrid classes, offered for credit in Fall 2021 at the same institution. Following Rovai's original 20-item CCS, the 8-item CCS-SF yields two interpretable factors, connectedness and learning. This study confirms the two-factor structure of the CCS-SF, and concludes that it is a valid measure of classroom community among undergraduate students enrolled in remote, hybrid, and in-person introductory mathematics and statistics courses.

## Introduction

Introductory undergraduate mathematics and statistics courses play a pivotal role in STEM retention, and students' sense of belonging in these courses is a critical factor in their persistence (Ellis et al., 2014; Olson & Riordan, 2012; Seymour, 1997; Seymour et al., 2019; Tinto, 1993). As articulated by Rovai, sense of belonging is a crucial aspect of classroom community, which he defines as being comprised both of students' sense of connectedness to one another as well as their shared commitment to and responsibility for learning (2002a). Beyond persistence in STEM, correlations have also been found between students' sense of community and their academic motivation and success (McKinney et al., 2006) as well as between their sense of community and perceived learning (Rovai 2002b; Trespalacios & Perkins, 2016). As the workforce demand for STEM majors continues to increase (Olson & Riordan, 2012), an improved understanding of classroom community in introductory undergraduate mathematics and statistics is crucial. Classroom community plays a particularly important role in online learning, where physical separation may contribute to feelings of disconnectedness (Green et al., 2017; Richardson et al., 2017). To develop classroom community, it is essential to measure it efficiently and reliably.

---

[1] Department of Statistical Science, Duke University
[2] Department of Mathematics, Duke University
[3] Learning Innovation, Duke University

Rovai's 20-item Classroom Community Scale (CCS) is one of the most widely used tools for measuring classroom community among students at colleges and universities (Dellasega, 2021). The CCS was initially developed and validated for a population of online students in graduate courses at a U.S. institution (Rovai, 2002a) and continues to be used primarily in this setting (Beeson et al., 2019; Rovai, 2002b; Trespalacios & Perkins, 2016; Wiest, 2015; Shackelford & Maxwell, 2012). However, the CCS has also been used widely to measure community among undergraduate students (Kocdar et al., 2018; Robinson et al., 2019; Young & Bruce, 2011) and those studying in-person (Brown & Pederson, 2020; Rovai & Jordan, 2004), as well as in multiple settings outside of the U.S., including Australia (Holland & Pithers, 2012), China (Zhang et al., 2011), Iran (Ahmady et al., 2018), and Turkey (Erden Aydin & Gumus, 2016).

Despite its widespread use, the CCS has received criticism for its low construct validity (Barnard-Brak & Shiu, 2010; Cho & Demmans Epp, 2019), and at 20 items may contribute to respondent fatigue. Cho & Demmans Epp developed the 8-item Classroom Community Scale Short Form (CCS-SF) to address these concerns and validated it with a population of online students enrolled in MOOCS and graduate courses (2019). While other instruments for measuring sense of community, particularly in online courses, exist (see, for example, a summary in Table 1 on p.56 of Randolph & Crawford, 2013), none are as concise as the CCS-SF, which also has the advantage of being adapted from one of the most frequently used scales. Further use of the CCS-SF has been sparse, albeit inclusive of international student populations at both the graduate (Demmans Epp et al., 2020) and undergraduate level (Akatsuka, 2020). However, there have been no uses of the instrument for students studying in non-online settings, and no additional efforts to validate the instrument beyond its introduction (Cho & Demmans Epp, 2019).

This study aims to fill this void, validating the CCS-SF for use with undergraduate students, specifically those in introductory mathematics and statistics courses at a private U.S. institution. The massive shift of instruction online in the wake of the COVID-19 pandemic provided a fruitful setting in which to evaluate the use of the CCS-SF for this population. Having validated the instrument for undergraduate students studying online, further confirmatory analysis was then conducted to validate the CCS-SF for students enrolled in in-person and hybrid classes in Fall 2021. This is a first step towards validating the CCS-SF for a broader population, thereby providing researchers and practitioners with a concise and validated tool for assessing classroom community in a variety of post-secondary settings.

# Background

**Classroom Community**

We follow Rovai's definition of classroom community, which in turn draws strongly upon the theory of community developed by McMillan & Chavis (1986). The latter emphasizes the importance of both membership and acceptance as well as of shared beliefs and objectives (McMillan & Chavis, 1986). Specifying this definition to the classroom setting, Rovai theorizes that classroom community consists of two factors, connectedness and learning:

> Connectedness represents the feelings of students regarding their cohesion, spirit, trust, and interdependence. Learning represents the feelings of students regarding the quality of their construction of understanding and the degree to which they share values and beliefs concerning the extent to which their learning goals and expectations are being satisfied. (2002b, p.325)

That is, students with a strong sense of classroom community "possess a shared faith that members' educational needs will be met through their commitment to shared learning goals" (Rovai, 2002a, p.198). Equivalently, one might think of classroom community as students' affiliation and affinity in the context and service of learning, or as Young & Bruce write, "connections among students and between students and instructors that lead to increased learning" (2011, p.218).

**The Classroom Community Scale (CCS)**

The Classroom Community Scale (CCS) measures classroom community via self-report, asking students to reply to 20 items on a five-point Likert-type response scale of *strongly disagree, disagree, neutral, agree*, and *strongly agree* (Rovai, 2002a). 10 items comprise the connectedness component (e.g., "I feel that students in this course care about each other"), and the other 10 items relate to the learning environment (e.g., "I feel that I am given ample opportunities to learn"). Each response is assigned a value of 0 through 4, with the 10 negatively worded items (e.g., "I feel uneasy exposing gaps in my understanding") reverse-scored so that the most favorable responses are always assigned a value of 4, and the least favorable a value of 0. The total score on the scale is then computed by adding up the scores on each of the 20 items, with a score of 80 reflecting the strongest possible sense of community and a score of 0 the weakest. Similarly, the scores on the connectedness and learning subscales each vary between 0 and 40.

Rovai developed the CCS by first creating an initial set of 40 candidate items, which were then assessed for content validity by a panel of 3 university instructors of educational psychology and subjected to a preliminary factor analysis (2002a, p.201). Each item which was not unanimously rated by the panelists as "*totally relevant* to sense of community in a classroom setting" was eliminated from the scale, as was each item which did not account for salient loadings (at least 0.3) onto the two interpretable factors of connectedness and learning (p.201). The result of these eliminations is the 20-item CCS, which was then evaluated by gathering data via optional questionnaire from 375 students enrolled in 28 online graduate courses offered for credit by a private university, a response rate of 66% (p.200). The two-factor structure of the scale was confirmed via the scree plot, the Kaiser-Gutman Rule, and solution interpretability via oblimin rotation, with the identified factors corresponding to the components of connectedness and learning (p.205). Reliability was confirmed via two internal consistency estimates, Cronbach's $\alpha$ and the split-half coefficient corrected by the Spearman-Brown prophecy formula. Both estimates indicated excellent reliability for the full scale (Cronbach's $\alpha$ and the equal-length split-half coefficient were 0.93 and 0.91, respectively), excellent reliability for the connectedness subscale (0.92 and 0.92, respectively) and good reliability for the learning subscale (0.87 and 0.80) (p.206).

**The Classroom Community Scale Short Form (CCS-SF)**

Cho & Demmans Epp's Classroom Community Scale Short Form (CCS-SF), shown in Table 1, consists of 8 of the 20 items from the original CCS, with identical wording and scoring. Whereas the CCS connectedness and learning subscales each comprised 10 items, the CCS-SF subscales each comprise 4 items. (Note that only 3 of the items in the CCS-SF are negatively worded, and all are in the learning subscale.) The total score on the scale varies between 0 and 32, with scores on the subscales each varying between 0 and 16, with a higher score again indicating a stronger sense of classroom community.

To develop the CCS-SF, Cho & Demmans Epp used Cronbach's $\alpha$ and exploratory factor analysis to investigate student responses to the CCS, gathering data via optional questionnaire from 197 students enrolled in 9 online Massive Open Online Courses (MOOCs) or graduate courses (response rate and institution information are not reported) (2019, p.2). First, the two-factor structure of the CCS was confirmed, with factors corresponding to the connectedness and learning subscales. A raw $\alpha$ of 0.80 was found, indicating good reliability, but four factor loading issues arose (of factors either not loading onto either factor, loading onto the incorrect factor, or loading onto both factors), calling into question the construct validity of the instrument. (As in the development of the original CCS, a threshold of 0.3 was used to indicate salient loading).

**Table 1**

*Items in Classroom Community Scale Short Form (CCS-SF)*

| Item | CCS # | Subscale | Question |
| --- | --- | --- | --- |
| 1 | 1 | Connectedness | I feel that students in this course care about each other. |
| 2 | 3 | Connectedness | I feel connected to others in this course. |
| 3 | 4 | Learning | I feel it is hard to get help when I have a question. |
| 4 | 8 | Learning | I feel uneasy exposing gaps in my understanding. |
| 5 | 10 | Learning | I feel reluctant to speak openly. |
| 6 | 13 | Connectedness | I feel that I can rely on others in this course. |
| 7 | 16 | Learning | I feel that I am given ample opportunities to learn. |
| 8 | 19 | Connectedness | I feel confident that others will support me. |

*Note.* Numbers in the second column indicate item numbers in the original CCS.

Starting with the full 20-item set, items were iteratively dropped from the scale to improve reliability (via improved $\alpha$) and validity (via factor loading) at each step. (Following the procedure used in the development of the original CCS, factor loading was investigated via maximum likelihood exploratory factor analysis using oblimin rotation.) The result of this process was a reduced 15-item scale (8 items on the connectedness subscale and 7 on the learning subscale), achieving an $\alpha$ of 0.82, an improvement of 0.02 from the initial set, but with three factor loading issues. To address these issues, a second construct validity investigation was conducted, now prioritizing dropping items with improper loading over dropping items to improve $\alpha$. This process resulted in the final 8-item CCS-SF, achieving an $\alpha$ of 0.74 (a decline of -0.06 from the initial set, but still within bounds of acceptability) with no factor loading issues.

At less than half the length of the original CCS with evidence of improved construct validity, the CCS-SF is a compelling instrument in need of further investigation. We undertake this work here.

# Methods

### Data and Participants

The primary data for this research were collected from 351 students in 21 introductory-level mathematics and statistics classes (one offered through the Economics department) during the Fall 2020 and Spring 2021 semesters. Additional data were collected from 128 students in 13 introductory-level mathematics and statistics classes during Fall 2021. All enrolled students in the included courses were invited to complete an optional questionnaire that included the CCS-SF items. The survey also asked students to identify the format of their course (online, in-person, or hybrid) and how they most often attended class (in-person or online), as well as self-reported demographics.

The researchers gained ethics approval from the University's Campus Institutional Review Board before the commencement of the study (Protocol 2021-0050). The analysis did not begin until after the conclusion of the semesters during which data were collected.

### Validation & Reliability Methods

We began by assessing the reliability of the instrument for an online undergraduate student population using Cronbach's $\alpha$. We then validated the instrument for this population using an approach similar to that used by Cho & Demmans Epp in their development of the CSS-SF (2019). We used confirmatory factor analysis to determine if the results from the data mapped onto the two factors associated with connectedness and learning as specified in Table 1. Several criteria were considered in this assessment, including the Comparative Fit Index (CFI), Tucker-Lewis Index (TLI), and the Root Mean Square Error of Approximation (RMSEA). We then assessed the optimal number of factors based on several criteria - Very Simple Structure, Velicer Minimum Average Partial, Empirical Bayesian Information Criterion (eBIC), and Bayesian Information Criterion (BIC). Finally, exploratory factor analysis was used to evaluate whether the factor loadings were consistent with the two subscales of connectedness and learning, and to assess the loadings in scenarios of three or more factors. All analysis was conducted using the *psych* R package (Revelle, 2015).

# Results

### Survey Response Rates

A total of 351 students completed the survey across the Fall 2020 and Spring 2021 semesters. This represents an average response rate of 17.49%, with class-level response rates varying between 4.49% and 60.47%. The low response rate can be attributed to two primary factors. First, students were recruited for this survey only via email. Response rates tend to be lower for surveys in which recruitment is done via email (Sappleton & Lourenço, 2016). One meta-analysis found that response rates for emailed surveys were an average of 20% lower than mailed surveys (Shih & Fan, 2009). In addition, students were surveyed towards the end of the semester when they were also receiving several other survey solicitations. It is likely that students were experiencing email overload and survey fatigue which led some of them to disregard the recruitment email (Dabbish & Kraut, 2006).

**Demographics of Survey Respondents**

Table 2 shows the demographic breakdown for students who participated in the surveys across the three semesters. In each semester, a plurality of the respondents were White non-Hispanic, a majority were female, and a majority were first-year students. Due to restrictions in data availability, we can make only limited comparisons of the demographics of survey respondents to the demographics of all students in the participating courses. In particular, we only have access to demographic data for students enrolled in introductory math courses (not statistics courses) in Fall 2019 and Spring 2020, the academic year right before our data collection period. Based on these data, about 54.8% of students in introductory math courses were female (Akin & Viel, 2022). Thus, females may be overrepresented among our survey respondents. About 53.3% of students in introductory math courses in Fall 2019 and Spring 2020 were White non-Hispanic (Akin & Viel, 2022), thus this group of students may be slightly underrepresented among our survey respondents. These differences in the distribution of race and sex between the survey respondents and the introductory math population of reference could be partially attributable to the distributions by race and sex in introductory statistics courses, as well as to the variability in student identities between semesters.

The introductory math and statistics courses at the institution where data were gathered often have a specific number of spots reserved for first-year students, and the percentage of first-year students among respondents is consistent with what we would expect given the introductory nature of the course and the reserved enrollment slots. Additionally, there is often a higher percentage of first-year students taking these courses in the Fall semesters, which is also reflected among our survey respondents.

**Table 2**

*Demographic Description of Survey Respondents*

| Demographic factor | Fall '20 (N = 229) | Spring '21 (N = 122) | Fall '21 (N = 128) |
|---|---|---|---|
| **Race** | | | |
| Hispanic | 22 (9.6%) | 14 (11%) | 20 (16%) |
| White non-Hispanic | 103 (45%) | 53 (43%) | 45 (35%) |
| Black non-Hispanic | 23 (10%) | 8 (6.6%) | 8 (6.2%) |
| Asian non-Hispanic | 59 (26%) | 35 (29%) | 35 (27%) |

| Demographic factor | Fall '20 (N = 229) | Spring '21 (N = 122) | Fall '21 (N = 128) |
|---|---|---|---|
| Other non-Hispanic | 22 (9.6%) | 12 (9.8%) | 20 (16%) |
| **Sex** | | | |
| Female | 141 (62%) | 77 (63%) | 92 (72%) |
| Male | 88 (38%) | 45 (37%) | 36 (28%) |
| **Year** | | | |
| First-year | 145 (63%) | 71 (58%) | 95 (74%) |
| Second – Fourth year | 84 (37%) | 51 (42%) | 33 (26%) |

**Reliability**

The Cronbach's $\alpha$ for the full data is 0.83. It is 0.84 for the connectedness subscale and 0.75 for the learning subscale. All of these values exceed the standard threshold of 0.7, as well as the value of 0.74 for the full scale obtained by Cho & Demmans Epp (2019). This indicates high internal consistency and reliability of the survey instrument.

**Confirmatory Factor Analysis**

The data map onto the two factors associated with connectedness and learning as specified in Table 1. The CFI is 0.87 and the TLI is 0.8. The RMSEA is 0.15 (95% CI: 0.13, 0.17). Note that the RMSEA is greater than the standard threshold of 0.1; however, taking all criteria into account, the two factors are confirmed.

**Exploratory Factor Analysis**

Very Simple Structure (VSS), Velicer Minimum Average Partial (VMAP), Empirical Bayesian Information Criterion (eBIC), and Bayesian Information Criterion (BIC) were used to determine the optimal number of factors for the data. Several rotations were considered, and, given the high correlation between the factors of interest and in agreement with the previous work in development of the CCS (Rovai, 2002a) and CCS-SF (Cho & Demmans Epp, 2019) the oblimin rotation was determined to be the one that best fit the data. This rotation was used throughout the exploratory data analysis.

**Table 3**

*Criteria for Exploratory Factor Analysis*

| Factors | VSS Fit | VMAP | eBIC | BIC |
|---|---|---|---|---|
| 1 | 0.000 | 0.068 | 175.848 | 190.676 |
| 2 | 0.888 | 0.063* | -41.612* | -10.055 |
| 3 | 0.896* | 0.101 | -29.423 | -8.164 |
| 4 | 0.885 | 0.148 | -11.311 | -10.609* |

*Note.* The optimized value for each criterion is indicated by an asterisk.

The criteria in Table 3 disagree about the optimal number of factors for this data. Based on VSS, the optimal number of factors is three; however, four is the optimal number of factors based on BIC, and VMAP and eBIC agree that two is the optimal number. The choice of two factors based on VMAP and eBIC is in alignment with the initial development of the CCS-SF instrument (Cho & Demmans Epp, 2019), which in turn confirmed the structure of the original CCS (Rovai, 2002a).

Most of the variability in the data is explained by the first two factors (53.2% and 17.34% respectively). The third factor explains an additional 10.48% of the variability, and the fourth factor explains about 8.62%.

Given the disagreement between the criteria in Table 3, we consider the results using two, three, and four factors. As in the analyses accompanying the development of the CCS-SF (Cho & Demmans Epp, 2019) and its predecessor the CCS (Rovai, 2002a), we required that an item has a loading of at least 0.3 to be considered loaded onto a factor.

### Two Factors

The two-factor loadings are shown in Table 4. Factor 1 is associated with connectedness, as the items in the connectedness subscale in Table 1 load onto this factor. One learning item, "I feel that I am given ample opportunities to learn" (Item 7) also loads onto Factor 1. The remaining items load onto Factor 2, associated with learning.

**Table 4**

*Factor Loadings for Two Factors*

| Item | Subscale | Connectedness Factor | Learning Factor |
|---|---|---|---|
| 1 | Connectedness | **0.700** | -0.011 |
| 2 | Connectedness | **0.743** | -0.015 |
| 3 | Learning | 0.202 | **0.409** |
| 4 | Learning | -0.045 | **0.940** |
| 5 | Learning | 0.101 | **0.699** |
| 6 | Connectedness | **0.770** | -0.119 |
| 7 | Learning | **0.513** | 0.222 |
| 8 | Connectedness | **0.805** | 0.088 |

*Note.* Loadings in bold text indicate the item loaded onto that factor.

### Three Factors

Table 5 contains the factor loadings for three factors. Similar to the two-factor result, Factor 1 is again associated with connectedness, with all of the items in the connectedness subscale loading onto this factor, along with Item 7 from the learning subscale which once more loads incorrectly. However, the addition of the third factor provides a more nuanced view of the learning subscale, now broken down into Factor 2 and Factor 3. The items "I feel uneasy exposing gaps in my understanding" (Item 4) and "I feel reluctant to speak openly" (Item 5) strongly load onto Factor 2, and the item "I feel it is hard to get help when I have a question" (Item 3) strongly loads onto Factor 3.

**Table 5**

*Factor Loadings for Three Factors*

| Item | Subscale | Connectedness Factor | Learning Factor (a) | Learning Factor (b) |
|---|---|---|---|---|
| 1 | Connectedness | **0.72** | 0.06 | -0.11 |
| 2 | Connectedness | **0.73** | -0.04 | 0.06 |
| 3 | Learning | 0.01 | 0.01 | **0.99** |
| 4 | Learning | -0.04 | **0.83** | 0.07 |
| 5 | Learning | 0.05 | **0.84** | -0.06 |
| 6 | Connectedness | **0.77** | -0.10 | -0.02 |
| 7 | Learning | **0.47** | 0.10 | 0.22 |
| 8 | Connectedness | **0.79** | 0.08 | 0.04 |

*Note.* Loadings in bold text indicate the item loaded onto that factor.

### Four Factors

The four-factor loadings in Table 6 repeat the refinement of the learning subscale seen in the three-factor result. In particular, Factor 2 in Table 5 is similar to Factor 2 in Table 4, as it includes the items regarding reluctance to speak (Item 5) and feeling uneasy exposing gaps (Item 4), and

Factor 4 in Table 5 is similar to Factor 3 in Table 4. However, the addition of a fourth factor now also provides a more nuanced view of the connectedness subscale. The items "I feel confident that others will support me" (Item 8) and "I feel that I can rely on others in this course" (Item 6) load onto Factor 1. The items "I feel connected to others in this course" (Item 2) and "I feel that students in this course care about each other" (Item 1) load onto Factor 3.

As with the results in Tables 4 and 5, these four factors align with the connectedness and learning subscales in Table 1, with the exception of Item 7 associated with finding ample opportunities to learn. This item does not load onto any factors based on the threshold of 0.3.

**Table 6**

*Factor Loadings for Four Factors*

| Item | Subscale | Connectedness Factor (a) | Learning Factor (a) | Connectedness Factor (b) | Learning Factor (b) |
|---|---|---|---|---|---|
| 1 | Connectedness | 0.18 | 0.10 | **0.58** | -0.11 |
| 2 | Connectedness | -0.04 | 0.00 | **0.86** | 0.06 |
| 3 | Learning | 0.01 | 0.01 | 0.00 | **0.99** |
| 4 | Learning | -0.05 | **0.88** | 0.00 | 0.05 |
| 5 | Learning | 0.07 | **0.79** | 0.00 | -0.05 |
| 6 | Connectedness | **0.47** | -0.08 | 0.34 | -0.02 |
| 7 | Learning | 0.26 | 0.11 | 0.24 | 0.22 |
| 8 | Connectedness | **0.99** | 0.02 | -0.01 | 0.02 |

*Note.* Loadings in bold text indicate the item loaded onto that factor.

### Determining the Number of Factors

After examining the analysis for two, three, and four factors, the two-factor loadings in Table 4 are chosen. Similar to the findings of Cho & Demmans Epp (2019), these two factors map onto the subscales of connectedness and learning, with the exception of Item 7. Additionally, the third and fourth factors did not explain much variability. In future iterations of the project, statistical modeling will be used to understand the association between student characteristics and pedagogy and sense of community. Preliminary exploration of the models indicated that two factors were most feasible for this purpose, as there is evidence of potential collinearity issues in models with three or four factors.

### Testing on Fall 2021 Data

The two-factor model in Table 4 is applied to data collected from 128 students in 13 introductory math and statistics courses in Fall 2021. During this semester, students participated in in-person or hybrid classes, with no courses being offered fully online. The Cronbach's $\alpha$ for the full Fall

2021 data is 0.82. For the connectedness subscale, $\alpha$ = 0.81, and for the learning subscale, $\alpha$ = 0.73. These values indicate high internal consistency and thus reliability of the survey instrument in this validation semester.

In the confirmatory factor analysis, CFI is 0.87, the TLI is 0.81, and RMSEA is 0.13 (95% CI:0.10, 0.17). The CFI and TLI exceed the threshold. RSMEA is still higher than the desired threshold of 0.1 or lower, but the two factors are a reasonable fit for this data.

# Discussion

Based on Cronbach's $\alpha$ and the confirmatory and exploratory factor analysis, the Classroom Community Scale Short-Form (CCS-SF,Cho & Demmans Epp, 2019) has been validated for the population of undergraduate students in in-person, online, and hybrid introductory math and statistics courses. The analysis showed the survey items mapped onto the two subscales measuring connectedness and learning, with high internal reliability and internal consistency.

**Contribution to the Field**

Much recent attention has been paid to classroom community in undergraduate STEM, particularly with the massive shift to online learning in Spring 2020 that accompanied the onset of the Covid-19 pandemic (see, e.g., McGee & Tashakkori, 2021). Literature has found many positive associations with a strong sense of community, including perceived learning (Rovai 2002b; Trespalacios & Perkins, 2016), academic motivation and success (McKinney et al., 2006), and, specifically with regards to the connectedness aspect, persistence in the field (Ellis et al., 2014; Olson & Riordan, 2012; Seymour, 1997; Seymour et al., 2019; Tinto, 1993).

By validating the 8-item CCS-SF (Cho & Demmans Epp, 2019), which was previously unvalidated for undergraduate students, this paper makes available to researchers a very concise tool for measuring classroom community in introductory quantitative science. The CCS (Rovai, 2002a) from which the CCS-SF is adapted is already the shortest among the 3 most widely used instruments for measuring connectedness among online students (Delasega, 2021) at 20 items. The other two instruments, the Community of Inquiry Survey Instrument (Arbaugh et al., 2008), and the Online Student Connectedness Survey (Bolliger & Inan, 2012), are 34 items and 25 items, respectively. In addition, this paper makes the unique contribution of validating the CCS-SF not only for undergraduate students in online courses but for those enrolled in hybrid and in-person offerings as well. While other instruments exist for measuring students' sense of belonging in mathematics, including the seminal Sense of Belonging to Math scale by Good et al. (2012), these instruments are also longer, and focus on the connectedness aspect of classroom community, whereas the CCS-SF also encompasses students' shared sense of responsibility for learning.

**Factor Loading Issue**

In the two-factor exploratory factor analysis (see Table 4), Item 7 regarding ample opportunities to learn mapped onto the connectedness subscale, though it is specifically asking about the learning environment. One potential reason for this is the wording of the item. All the items evaluating connectedness are positively worded, such that stronger agreement indicates stronger sense of community. However, with the exception of Item 7, the items in the learning subscale are negatively worded, such that stronger agreement indicates a lesser sense of perceiving the learning environment as supportive. Because Item 7 is the only item in the learning subscale that is positively worded, students may have associated it with connectedness as they answered the items. This is consistent with conclusions from Good et al. (2012) about different loadings for positively and negatively worded items even if they are capturing the same phenomenon. Though this item loaded onto a different factor than it did in Cho & Demmans Epp (2019), the validity of this item's mapping was shown here for the Fall 2021 student population.

**Insights From Four-Factor Loadings**

While overall analysis confirmed the two-factor structure of the instrument, insights can be gleaned about the nuances of student experiences of connectedness and learning from the four-factor loadings in Table 6. In particular, Factors 2 and 4 refine the learning subscale, and Factors 1 and 3 refine the connectedness subscale.

The *classroom climate*, otherwise known as *learning environment*, is "intellectual, social, emotional, and physical environments in which our students learn" (Ambrose et al., 2010, p.170). Students' perceptions of a more positive learning environment are associated to higher perceptions of learning and higher academic achievement (Ertmer and Stephich, 2005; Harvey et al., 2007; Wighting et al., 2009). Factors 2 and 4 in Table 6 give more granular insight into students' perceptions of different components of the learning environment. Factor 2 addresses students' reluctance to speak up in learning environments, particularly if this has the potential to expose potential gaps in their understanding of the course content. Similarly, Factor 4 more directly incorporates student interactions and feelings toward the instructional team in addition to their peers. Given the impact that learning environment has on students' outcomes, this more granular view of its components can enable researchers and educators to identify specific interventions to help create a more positive and inclusive learning environment.

Complementing the classroom climate is students' sense of connectedness with their peers. In higher educational research, "connectedness" is often used synonymously with "belonging" (Lahdenperä & Nieminen, 2020). Much work has been done to examine students' sense of belonging in STEM fields at large (see, e.g., Wilson et al., 2015; Jackson 2016; Rainey et al., 2018), and more specifically in quantitative science (see, e.g., Lahdenperä & Nieminen, 2020; Moudgalya et al., 2021), which frequently uses Good et al.'s well-known Sense of Belonging to Math scale (2012). This scale yielded five interpretable factors: Membership, Acceptance, Affect, Trust, and Desire to Fade. The refinement of the Connectedness subscale into Factor 1 and Factor 3 could be viewed as a coarsening of Good et al.'s factors. In particular, Factor 1 primarily encompasses the degree to which students feel they are supported by their classmates and to a lesser degree whether they feel they can rely on their classmates if needed. This could be viewed as a coarsening of Acceptance and Desire to Fade. Similarly, Factor 3 consists of the degree to which students feel connected to their classmates and whether they feel their classmates care about them. This could be viewed as a coarsening of Good et al.'s factors of Membership and Affect. Good et al.'s Trust factor focuses on the role of the instructor, and is more closely aligned with the Learning subscale of the CCS and CCS-SF.

**Limitations**

There are some limitations to this work that should be considered. The first is that most of the data collection occurred in the Fall 2020 and Spring 2021 semesters, still in the height of the COVID-19 pandemic with substantial disruptions to education. Students' sense of connectedness and learning in their courses could have been impacted, given the unprecedented societal and personal challenges faced by so many during that time. This in part includes students at this institution being required to take courses online rather than self-selecting into online and hybrid courses. We mitigate some of this concern by validating the instrument on the Fall 2021 data once courses at this institution were largely back in-person.

The other major limitation to the study is the relatively low response rate for the surveys. Because data were collected over multiple semesters, we were able to obtain a reasonable sample size for conducting these analyses; however, we would ideally have responses from more students in these courses. Because the survey data is all anonymized in the analysis and no overall demographic course data is available, we are unable to definitively identify if there are fundamental differences between the students who completed the survey and those who did not. Given the increased number of surveys administered to students and the aforementioned unprecedented circumstances, the level of response received in these surveys is consistent with other studies at the institution conducted in the similar time.

A final caveat is the setting for the study: all data were drawn from students enrolled at the same private, highly selective United States institution. However, as shown in Table 2, the respondents were diverse along racial and gender axes. Given previous use of the CCS-SF in international (Akatsuka, 2020) and public (Demmans Epp et al., 2020) university settings, we anticipate that researchers at other types of institutions will be able to use the methods established in this study to investigate the validity of the CCS-SF for a broader population.

## Conclusion and Future Work

The primary objective of this research project is to understand classroom practices that are associated with an increased sense of classroom community among students in introductory undergraduate math and statistics courses. This article has established the validity of the Classroom Community Scale Short-Form (CCS-SF) instrument (Cho & Demmans Epp, 2019) for measuring classroom community among this population in a variety of class formats, confirming the two-factor structure of connectedness and learning.

Therefore, the immediate next steps are to use statistical modeling to analyze how classroom practices and policies specified by the instructors of these introductory courses are associated with connectedness and learning, and whether these relationships differ for students with different identities. We will also conduct focus groups with students from these courses to provide more context and detail about some of the phenomena revealed in the analysis of the survey data. Long term, we intend to extend this research to other institutions in the U.S. and beyond, so we can better understand sense of community at a range of institutions thus more fully capturing the diversity of the global student population in introductory undergraduate math and statistics.